\documentclass [12 pt]{article}
\usepackage{amsmath}
\usepackage{multirow}
\usepackage{caption}

\begin{document}

\title{Hypothesis Testing of Blip Effects in Sequential Causal Inference}
\author{Xiaoqin Wang$^1$ and Li Yin$^2$}

 \date{}
\maketitle


\noindent
$^1$University of G\"{a}vle,  Sweden.

\noindent
$^2$Karolinska Institutet,  Sweden. Correspondence: Li.Yin@ki.se



\begin{abstract}
In this article, we study the hypothesis testing of the blip / net effects of   treatments in a treatment sequence.
We illustrate  that the likelihood ratio test and the score test may suffer from  the curse of dimensionality,  the null paradox  and the high-dimensional constraint on  standard parameters under the null hypothesis. On the other hand, we construct the Wald   test via a small number of  point effects of treatments in  single-point causal inference.  We show that the  Wald   test can avoid these problems under the same assumptions as  the Wald test for testing the point effect of  treatment. The simulation study illustrates that the Wald test achieves the nominal level of  type I error and a  low level of type II error. A real medical example illustrates how to conduct the Wald test in practice.
\end{abstract}


{\noindent {Mathematics Subject Classification (2000):  62F03;  62F05; 62F30}

\newtheorem{E}{Example}
\newtheorem{T}{Theorem}
\newtheorem{PP}{Proposition}
\newtheorem{C}{Corollary}
\newtheorem{D}{Definition}
\newtheorem{R}{Remark}
\newtheorem{LL}{Lemma}
\newtheorem{A}{Assumption}

\section{Introduction}
In many economic and medical researches,  treatments are assigned in the form of a sequence to influence an outcome of interest that occurs after the last treatment of the sequence. Between   treatments,  there are often  time-dependent covariates that   may be posttreatment variables of the earlier treatments  and confounders of the subsequent  treatments. The blip effect of treatment   is the causal effect of the treatment  on the    outcome given  the history of previous    treatments and covariates  while  setting the subsequent treatments at controls (Robins, 1997; Hernan and Robins, 2018). It is also called the net effect of treatment (Wang and Yin, 2015). The blip effect   plays a central role in sequential causal inference for the following reasons. The blip effect reveals   the effect modification of covariates on the outcome, which is of considerable interest in practice (Robin, 1997; Almirall et al., 2010;  Hernan and Robins, 2018).
From the blip effect of  treatment at a particular time, we can find the optimal  treatment at that   time  given the previous  treatments and covariates (Robin, 1997; Hernan and Robins, 2018). From the blip effects of all treatments in the sequence, we can obtain the causal  effect of any treatment regime on the outcome and find  the optimal treatment regime (Robin, 1997; Hernan and Robins, 2018; Wang and Yin, 2019).

When estimating the blip effects via standard parameters, two problems may occur: the curse of dimensionality and the null paradox.  The curse of dimensionality implies that  if a treatment sequence is  long and / or the number of covariates is large, a huge number of standard parameters are needed  in the  estimation.  The null paradox implies that even  if the blip effects are all null, any unsaturated model  is  misspecified that imposes equalities between standard  parameters.
Several methods are available to address the two problems, which include the parametric model based on the well-known $G$-formula (Taubman et al., 2009), the marginal structural model based on the inverse probability of treatment weighting (Robins, 1999) and the doubly robust method based on the structural nested mean model  (SNMM) (Robins, 1997).
SNMM describes a pattern of the blip effects and is  specified as a deterministic function indexed by a   parameter vector  of small dimension.

When testing the blip effects via standard parameters, additional problem may occur: the estimation needs to be carried out under the null hypothesis, which is expressed as   a constraint among standard parameters.
For a long treatment sequence and plentiful covariates, the constraint consists of many complex equations in standard parameters. This high-dimensional constraint implied by the null hypothesis,
together with the curse of dimensionality and the null paradox, makes the testing problem highly difficult.
Probably due to these three  problems, there are few literatures on the hypothesis testing of the blip effects. To the best of our knowledge, only
one attempt has been made in which  the authors used  the doubly robust method to model SNMM under special circumstances, where  strong assumptions   are imposed on SNMM and the method (Wallace et al., 2016).

On the other hand, Wang and Yin (2019) derived the new $G$-formula, which identifies the blip effect via the point effects of treatments. The point effect  is simply the point effect of   treatment in the framework of single-point causal inference, and its estimation and hypothesis testing have been well studied (Rosenbaum and Rubin, 1983). Hopefully,  the new $G$-formula will help to find a workable way of testing the blip effects.

In this article, we study the hypothesis testing of blip effects.
In Section $2$, we describe the hypothesis   and illustrate that  the likelihood ratio test and the score  test suffer from the curse of dimensionality, the null paradox and the high-dimensional constraint  on standard parameters under the null hypothesis. In Section $3$, we construct the Wald test via point effects  and show that it does not necessarily suffer from these problems. In Section $4$, we illustrate some finite sample properties of the Wald   test by simulation. In Section $5$, we illustrate the application of our method via a real medical example. In Section $6$, we conclude the article with discussion.

\section{Hypothesis testing  of blip effects  in sequential causal inference}
Here we will introduce key elements of this work in Section $2.1$ and   review the blip effect, the point effect and the new $G$-formula  in Section $2.2$. Then we will explicate the hypothesis on the blip effects in Section $2.3$ and finally
 illustrate with an example the problems from which the likelihood ratio test and the score test suffer.
\subsection{Treatments, covariates  and the outcome}
Consider a set of random variables in the temporal order: $(\mathbf X_1, Z_1, \mathbf X_2, Z_2, $ $\ldots,
 \mathbf X_{T}, Z_T, Y)$, where $Z_t$ are the treatment variables at times $t=1,\ldots,T$,
 $\mathbf X_1$ is a stationary covariate vector before $Z_1$, $\mathbf X_t$ ($t=2,\ldots, T$) are time-dependent covariate vectors between $Z_{t-1}$ and $Z_t$, and $Y$ is the outcome of interest after the last treatment $Z_T$.
Let  $\mathbf Z_1^t=(Z_1, \ldots, Z_t)$,
$\mathbf{X}_{1}^{t}$ $=$ $(\mathbf{X}_{1},
,\ldots, \mathbf{X}_{t})$, and $( \mathbf X_1^{T},\mathbf Z_1^T, Y)=$ $(\mathbf X_1, Z_1,  \ldots, \mathbf X_{T}, Z_T, Y)$. These variables  have the  realizations
 $(\mathbf x_1^{T},\mathbf z_1^T,   y)=(\mathbf x_1, z_1,  \ldots, \mathbf x_{T}, z_T, y)$.

In the following, we will use  ${\rm P}(.)$ to denote  the  probability  distribution of discrete   variables or the  density distribution of continuous  variables. The joint distribution of  $(\mathbf X_1^{T},\mathbf Z_1^T,  Y)$ is given by
\begin{equation}\label{eq1.0}
{\rm P}(\mathbf{x}_1^{T}, \mathbf z_1^T,y)
\end{equation}
$$
= {\rm P}(\mathbf x_1) {\rm P}(z_1   \mid \mathbf x_1)\cdots
{\rm P}(\mathbf x_T \mid \mathbf x_1^{T-1}, \mathbf z_1^{T-1})
{\rm P}(z_T \mid \mathbf x_1^{T}, \mathbf z_1^{T-1})
{\rm P}(y\mid \mathbf x_1^{T},  \mathbf z_1^{T}   ).
$$
The standard parameter for the conditional distribution ${\rm P}(\mathbf x_{t}\mid \mathbf x_1^{t-1},  \mathbf z_1^{t-1})$ of covariate  $\mathbf X_{t}$ given the history
$(\mathbf x_1^{t-1},  \mathbf z_1^{t-1})$ is the conditional mean $E(\mathbf X_{t} \mid \mathbf x_1^{t-1},  \mathbf z_1^{t-1})$. Let $\Psi_{\mathbf x_{t}}=\{E(\mathbf X_{t} \mid \mathbf x_1^{t-1},  \mathbf z_1^{t-1}) \}$ be the set of all these standard parameters for
${\rm P}(\mathbf x_{t}\mid \mathbf x_1^{t-1},  \mathbf z_1^{t-1})$ at $t$.
The standard parameter for ${\rm P}(\mathbf z_t\mid \mathbf x_1^{t},  \mathbf z_1^{t-1})$ is   $E(\mathbf Z_t \mid \mathbf x_1^{t},  \mathbf z_1^{t-1})$.
    Let $\Psi_{\mathbf z_t}=\{E(\mathbf Z_t \mid \mathbf x_1^{t},  \mathbf z_1^{t-1}) \}$. The standard parameter for  ${\rm P}(y\mid \mathbf x_1^{T},  \mathbf z_1^{T})$   is    $\mu(\mathbf x_1^{T},  \mathbf z_1^{T})=E(Y\mid \mathbf x_1^{T},  \mathbf z_1^{T})$. Let $\Psi_y=\{\mu(\mathbf x_1^{T},  \mathbf z_1^{T})\}$.

\subsection{Blip effects, point effects and the new $G$-formula}
Throughout the article, we assume
that there is no unmeasured confounder for the assignment of  treatment $z_t$ (Robins, 1997; Hernan and Robins, 2018).
Although the assumption is not testable, it can be satisfied in practice.
The assumption  is satisfied in sequential randomized experiments where  $z_t$ is randomly assigned according to a  history $(\mathbf x_1^{t},\mathbf z_1^{t-1})$ of the earlier   treatments and covariates. It is approximately  satisfied  in observational studies with a sufficient array $\mathbf x_1^{t}$ of covariates.

The
blip effect  $\phi( \mathbf x_1^{t}, \mathbf z_1^{t-1}; z_t)$    is the causal effect of active treatment  $z_t> 0$ relative to control treatment $z_t =0$
on the outcome $Y$ given $( \mathbf x_1^{t},\mathbf z_1^{t-1})$ when the subsequent treatments are set at controls, that is, $\mathbf z_{t+1}^T=\mathbf 0$. Under  the assumption of no unmeasured confounders, the blip effect can be expressed    in terms of the standard parameters
$\mu( \mathbf x_1^{T},\mathbf z_1^T)$ by
the well-known  $G$-formula (Robins, 1997; Hernan and Robins, 2018). It can also be expressed in terms of the point effects  by the new $G$-formula (Wang and Yin, 2019), as described below.

Let $\mu(\mathbf x_1^{t}, \mathbf z_1^{t})=E(Y\mid \mathbf x_1^{t}, \mathbf z_1^{t})$  be the conditional mean of $Y$ given $(\mathbf x_1^{t}, \mathbf z_1^{t})$.  The point  effect of treatment $z_t > 0$  is
\begin{equation}\label{eq10.0}
\theta(\mathbf{x}_{1}^{t}, \mathbf{z}_{1}^{t-1}; z_t)=\mu(\mathbf x_1^{t}, \mathbf z_1^{t-1},z_t)- \mu(\mathbf x_1^{t}, \mathbf z_1^{t-1},z_t=0).
\end{equation}
Noticeably, $\theta(\mathbf{x}_{1}^{t}, \mathbf{z}_{1}^{t-1}; z_t)$ is simply the point effect of   treatment $z_t$ in single-point causal inference, and its estimation and hypothesis testing have been well studied (Rosenbaum and Rubin, 1983).

Under the assumption of no unmeasured confounders, Wang and Yin (2019) decomposed the point effect $\theta(\mathbf x_1^{t},\mathbf
z_1^{t-1};z_t)$ of $z_t$ into the blip effects of $z_t$ and the subsequent treatments  $\mathbf{z}_{t+1}^{T}$ by
\begin{equation}\label{eq10.1}
\theta(\mathbf x_1^{t},\mathbf
z_1^{t-1};z_t)=\phi(\mathbf x_1^{t},\mathbf
z_1^{t-1};z_t)
\end{equation}
$$
+\sum_{s=t+1}^T E_1 \{\phi( \mathbf x_{t+1}^{s},\mathbf z_{t+1}^{s-1} ;z_s)\mid \mathbf x_1^{t},\mathbf
z_1^{t-1},z_t\}
$$
$$
-\sum_{s=t+1}^T E_2 \{\phi(\mathbf x_{t+1}^{s},\mathbf z_{t+1}^{s-1} ;z_s)\mid \mathbf x_1^{t},\mathbf
z_1^{t-1},z_t=0\},
$$
where the conditional expectation $E_1(.)$ is with respect to    ${\rm P}(\mathbf{x}_{t+1}^{s},\mathbf z_{t+1}^{s-1}, z_s\mid \mathbf x_1^{t},\mathbf
z_1^{t-1},z_t)$ and $E_2(.)$ to ${\rm P}( \mathbf{x}_{t+1}^{s},\mathbf z_{t+1}^{s-1},z_s\mid \mathbf x_1^{t},\mathbf
z_1^{t-1},z_t=0)$. Noticeably  at $t=T$, $\theta(\mathbf x_1^{T},\mathbf
z_1^{T-1};z_T)=\phi(\mathbf x_1^{T},\mathbf
z_1^{T-1};z_T)$.
 They also derived  its converse form in which  the blip effect is expressed in terms of the point effects and called the two formulas
 the new $G$-formula for the blip effect.

\subsection{Hypothesis  on blip effects}
Robins (Robins, 1997;  Hernan and Robins, 2018) pointed out that in most practices, the blip effects  follow a certain pattern   described   by  SNMM
$$
\phi(\mathbf x_1^{t}, \mathbf z_1^{t-1} ; z_t)=f(\mathbf x_1^{t}, \mathbf z_1^{t-1}, z_t; \boldsymbol\gamma), \quad t=1,\ldots, T,
$$
 where $f(.)$ is a deterministic function of $(\mathbf x_1^{t}, \mathbf z_1^{t-1}, z_t)$ indexed by a   parameter vector $\boldsymbol\gamma=(\gamma_1, \dots, \gamma_k)'$ of small dimension called the blip effect parameter.  For example, a simple dose-effect relationship of SNMM is given by  $\phi(\mathbf x_1^{t}, \mathbf z_1^{t-1} ; z_t)=\gamma z_t$, where $\gamma$ is a one-dimensional vector; if there is additionally the effect modification by $\mathbf x_{t}$, then SNMM is described by
$\phi(\mathbf x_1^{t}, \mathbf z_1^{t-1} ; z_t)=\gamma_{1} z_t+ \boldsymbol\gamma_{2}'z_t\mathbf x_{t}$, where $\boldsymbol\gamma=(\gamma_{1}, \boldsymbol\gamma_2')'$ and $\boldsymbol\gamma_2$  has  the same dimension as $\mathbf x_{t}$.

In this article, we focus on a linear SNMM, namely,
\begin{equation}\label{eq1.4}
\phi(\mathbf x_1^{t}, \mathbf z_1^{t-1} ; z_t)=\sum_{j=1}^k\gamma_j f_j(\mathbf x_1^{t}, \mathbf z_1^{t-1},z_t), \quad t=1,\ldots, T.
\end{equation}
However,  our  method of testing $\boldsymbol\gamma$ can be extended to the non-linear SNMM.

In sequential causal inference, it is essential  to test  the blip effects, or equivalently, the blip effect parameter $\boldsymbol\gamma$ due to (\ref{eq1.4}).
In the rest of the article, we test the hypotheses    of the  form
\begin{equation}\label{eq1.5}
H_0:  \mathbf H\boldsymbol\gamma - \boldsymbol\rho=\mathbf 0  \quad {\mbox {\rm against}} \quad H_1:   \mathbf H\boldsymbol\gamma -\boldsymbol\rho \neq \mathbf 0
\end{equation}
where  $\mathbf H$ is a $l\times k$ matrix with $l \leq k$ and $\boldsymbol\rho$ is a constant $l$-dimensional vector.

\subsection{Problems with  likelihood ratio test and score test}
According to the well-known $G$-formula (Robins, 1997; Hernan and Robins, 2018), the blip effects are functions of all standard parameters for the joint distribution (\ref{eq1.0}) of the  treatments,  covariates and outcome.
 Therefore,
the likelihood ratio test on $\boldsymbol\gamma$ requires estimating  all  standard   parameters     under a constraint  implied by the hypothesis $H_0$. The score test requires calculating  the score functions for all   standard parameters and evaluating  these functions at the estimates of these standard  parameters obtained under the null hypothesis $H_0$. In the following, we illustrate three major problems of the two tests by an example.

We consider a treatment sequence of  length $T=10$, in which covariates $X_{t}$ and treatments $Z_t$ are all   dichotomous.  Suppose the null hypothesis $H_0$ is such that all blip effects are equal to one another and  can be described by a blip effect parameter $\gamma$ of one dimension. When testing the hypothesis by the likelihood ratio test or the score test, we need to estimate
a huge amount of standard parameters:   $2^{20}=1048576$ standard parameters for the conditional distribution  ${\rm P}(y \mid \mathbf x_1^{10},  \mathbf z_1^{10})$, $2^{19}=524288$ for ${\rm P}(z_{10} \mid \mathbf x_1^{10},  \mathbf z_1^{9})$,  $2^{18}=26144$ for ${\rm P}(x_{10} \mid \mathbf x_1^{9},  \mathbf z_1^{9})$, and $2^{17}+\cdots+2=262142$ for ${\rm P}(\mathbf x_1^{9},  \mathbf z_1^{9})$ (namely, the curse of dimensionality).  Even under the null hypothesis,
 these standard parameters are  essentially all different, because  covariate $X_{t}$ ($t=2,\dots, 10$) is a posttreatment variable of the earlier treatments $\mathbf Z_1^{t-1}$ and confounders of the subsequent treatments $\mathbf Z_{t}^{10}$ (namely, the null paradox).
The null hypothesis implies   a constraint consisting of $699049$ equations  in these standard parameters (namely, the high-dimensional constraint on standard parameters   under $H_0$).

In the next section, we will show that the Wald test has the flexibility of allowing for estimating and testing  $\boldsymbol\gamma$ via  a small number of the point effects instead of standard parameters and thus  does not necessarily suffer from these three problems.

\section{Wald  test for blip effects}
First, we will construct  a model for the point effects indexed by  the blip effect parameter in Section $3.1$.  Second,  we will use the model to  estimate   the blip effect parameter conditional on all treatments and covariates in Section $3.2$. Third,  we will use the conditional estimate to obtain the marginal estimate of the blip effect parameter in Section $3.3$.   Then, we will use the asymptotic distribution of the marginal estimate to construct the Wald test  for  the blip effect parameter in Section $3.4$. Finally, we introduce the practical procedure for conducting the  hypothesis test.

\subsection{Model for point effects}
SNMM describes a pattern of the blip effects and often has a simple form in practice.  To fix the idea,
we assume that the blip effect of treatment  $z_t$ depends only on the last   covariate $\mathbf x_{t}$. For instance, the blip effect of a blood pressure drug ($z_t$) usually depends only on the latest blood pressure  and the  prognosis factors   ($\mathbf x_{t}$).
In this case, SNMM (\ref{eq1.4}) becomes
\begin{equation}\label{eq10.3}
\phi(\mathbf x_1^{t}, \mathbf z_1^{t-1} ; z_t)=\sum_{j=1}^k\gamma_j f_j(\mathbf x_{t}, z_t).
\end{equation}
Furthermore,
 the  assignment of $z_t$ often satisfies certain conditions  in practice, that is, it depends only on part of the history  $(\mathbf x_1^{t},\mathbf z_1^{t-1})$. Even in observation studies,
  the  assignment of $z_t$ can be   approximated
by a number of  sub randomized trials      called subclasses (Rosenbaum and Rubin, 1983).
  To fix the idea, we assume that it only depends on the latest covariate $\mathbf x_{t}$, so that
\begin{equation}\label{eq10.4}
{\rm P}(\mathbf x_1^{t-1}, \mathbf z_1^{t-1}\mid \mathbf x_{t}, z_t)={\rm P}(\mathbf x_1^{t-1}, \mathbf z_1^{t-1}\mid \mathbf x_{t}).
\end{equation}
We will use (\ref{eq10.3}) and (\ref{eq10.4}) to develop our testing method. However, our method can be applied to other SNMMs and treatment assignment conditions.

Consider the conditional mean
\begin{equation}\label{eq10.3.1}
 \mu( \mathbf x_{t},z_t)=E(Y\mid \mathbf
x_{t},z_t ),
\end{equation}
where the expectation is with respect to the conditional probability ${\rm P}(y\mid \mathbf
x_{t},z_t)$.
The  point   effect of $z_t > 0$ in stratum $\mathbf
x_{t}$ is
\begin{equation}\label{eq10.5}
\theta(\mathbf
x_{t}; z_t)=\mu(\mathbf
x_{t},z_t)-\mu(\mathbf
x_{t},z_t=0).
\end{equation}
As well-known in single-point causal inference (Rosenbaum and Rubin, 1983), formula (\ref{eq10.4}) implies
\begin{equation}\label{eq10.5.1}
\theta(\mathbf
x_{t}; z_t)=E\{\theta(\mathbf{x}_{1}^{t}, \mathbf{z}_{1}^{t-1}; z_t)\mid \mathbf
x_{t}, z_t\},
\end{equation}
where the expectation is with respect to ${\rm P}(\mathbf x_1^{t-1}, \mathbf z_1^{t-1}\mid \mathbf x_{t}, z_t)$.
In
Supplement I of Supporting Material, we also provide a proof for (\ref{eq10.5.1}).
Noticeably,  $\theta(\mathbf
x_{t}; z_t)$ are far fewer than   $\theta(\mathbf{x}_{1}^{t}, \mathbf{z}_{1}^{t-1}; z_t)$.

By decomposing the point effect $\theta(\mathbf x_{t-1};z_t)$ into  components $\gamma_j$ of the blip effect parameter $\boldsymbol\gamma$,
 we obtain the model for point effects
\begin{equation}\label{eq10.6.0}
\theta(\mathbf x_{t};z_t)=\sum_{j=1}^k \gamma_j c_j(\mathbf x_{t};z_t), \quad t=1,\ldots, T
\end{equation}
with
$$
c_j(\mathbf x_{t};z_t)=f_j(\mathbf x_{t},z_t)+\sum_{s=t+1}^T E_1 \{f_j( \mathbf x_{s},z_s)\mid \mathbf x_{t},z_t\}
$$
$$
-\sum_{s=t+1}^T E_2 \{f_j(\mathbf x_{s},z_s)\mid \mathbf x_{t},z_t=0\},
$$
where the expectation $E_1(.)$ is with respect to ${\rm P}(\mathbf{x}_{s}, z_s\mid \mathbf x_{t},z_t)$ and $E_2(.)$ to ${\rm P}( \mathbf{x}_{s},z_s\mid \mathbf x_{t},z_t=0)$. Noticeably at $t=T$, $c_j(\mathbf x_{T};z_T)=f_j(\mathbf x_{T},z_T)$.
The
 $c_j(\mathbf x_{t};z_t)$ is a sum of the contributions to component  $\gamma_j$ of $\boldsymbol\gamma$  from stratum $(\mathbf x_{t},z_t)$ versus $(\mathbf x_{t},z_t=0)$.
 Given all   treatments and covariates $(\mathbf{x}_1^{T}, \mathbf{z}_1^{T})$, model (\ref{eq10.6.0})  is  an unsaturated model  for the point effects $\theta(\mathbf x_{t};z_t)$ and indexed by a $k$-dimensional blip effect parameter  $\boldsymbol\gamma$.
  In
Supplement I of Supporting  Material, we will provide a  proof for (\ref{eq10.6.0})
 by
applying  (\ref{eq10.3}) and (\ref{eq10.4}) to the new $G$-formula (\ref{eq10.1}).

For convenience, we use
 $\boldsymbol\theta_t$ to denote
 a  vector of the  point effects $\theta(\mathbf
x_{t}; z_t)$ for different $(\mathbf x_{t}, z_t)$  at time $t$. Putting all  $\boldsymbol\theta_t$ together, we obtain the point effect vector $\boldsymbol\theta=(\boldsymbol\theta_1', \dots, \boldsymbol\theta_T')'$. Let ${\mathbf c}(\mathbf x_{t}, z_t)=$  $\{c_1(\mathbf x_{t}, z_t),\dots, $ ${ {c}}_k (\mathbf x_{t},z_t) \} $.
We use $\mathbf C_t$ to denote the matrix with row vectors  ${\mathbf c}(\mathbf x_{t}, z_t)$ for different $(\mathbf x_{t}, z_t)$  at time $t$; Putting all $\mathbf C_t$ together, we obtain the design matrix  $\mathbf C=(\mathbf C_1', \dots, \mathbf C_T')'$. Then, model (\ref{eq10.6.0}) can be written in the vector form as
\begin{equation}\label{eq10.6}
\boldsymbol\theta=\mathbf C \boldsymbol \gamma.
\end{equation}
Applying  this model,
we identify $\boldsymbol\gamma$ by
\begin{equation}\label{eq10.7}
\boldsymbol\gamma=({\mathbf C}' {\boldsymbol  \Sigma}^{-1} { \mathbf C})^{-1} {  \mathbf C}'  { \boldsymbol \Sigma}^{-1} {\boldsymbol\theta},
\end{equation}
where  $\boldsymbol \Sigma$ is a  positive definite  matrix, which can be arbitrarily chosen such that the matrix  $({\mathbf C}' {\boldsymbol  \Sigma}^{-1} { \mathbf C})$ is invertible.

Several statements can be made about model (\ref{eq10.6.0}) or equivalently (\ref{eq10.6}). First,  the  sizes of  the point effect vector $\boldsymbol\theta$   and  the design matrix $\mathbf C$ are only proportional  to  the length $T$ of  treatment sequence; potentially we may use  (\ref{eq10.6}) to estimate the blip effect parameter $\boldsymbol\gamma$ and  overcome the curse of dimensionality. Second,
given  $(\mathbf{x}_1^{T}, \mathbf{z}_1^{T})$ and thus $\mathbf C$,  model (\ref{eq10.6}) is an unsaturated model for   the point effect vector $\boldsymbol\theta$  and indexed by a $k$-dimensional   parameter vector $\boldsymbol\gamma=(\gamma_1,\dots, \gamma_k)'$; potentially we  may use (\ref{eq10.6}) to improve the estimation and overcome the null paradox.
Third,  the  $\boldsymbol\gamma$ is the model parameter in (\ref{eq10.6}); potentially we may use (\ref{eq10.6}) to estimate $\boldsymbol\gamma$ under the null hypothesis  $H_0$,    avoiding  the high-dimensional constraint on standard parameters under $H_0$.

Let us look at the  example of Section $2.4$, where the treatment sequence is $T=10$,    treatments and covariates are all dichotomous, and the null hypothesis $H_0$ is such that all blip effects are equal to one another and described by a one-dimensional  blip effect parameter $\gamma$.
Then the point effect vector $\boldsymbol\theta$ consists of only $20$ point effects $\theta(x_{t};z_t=1)$ ($t=1,\dots, 10$; $x_{t}=0,1$).    Applying  (\ref{eq10.6.0}) or (\ref{eq10.6}), the point effect $\theta(x_{t};z_t)$
decomposes into
$$
 \theta(x_{t};z_t)=c( x_{t};z_t) \gamma, \quad t=1,\dots, 10,
$$
where $c(x_{t};z_t)=1+\sum_{s=t+1}^{10} \{{\rm P}(z_s=1\mid x_{t}, z_t) - {\rm P}(z_s=1\mid x_{t}, z_t=0)\}$. Thus, the design matrix $\mathbf C$ becomes a column  vector of $20$ elements $c( x_{t};z_t=1) $. Furthermore,
the model  is  unsaturated for $\boldsymbol\theta$ and indexed  by only one parameter $\gamma$. Meanwhile, it is also the model under  $H_0$.
In contrast,
the likelihood ratio test and the score test still suffer from the curse of dimensionality, the null paradox and the high-dimensional constraint on standard parameters under $H_0$, because all the standard parameters  are still involved in these tests and the time-dependent covariates are still posttreatment variables as well as confounders.

\subsection{Conditional estimate of  blip effect parameter given all   treatments and covariates}
Suppose   a data  of observations  $(\mathbf x_{i1}^{T},\mathbf z_{i1}^{T}
,y_i)$,  $i=1,\ldots,n$ from  the
 random variables $(
\mathbf X_{i1}^{T},\mathbf Z_{i1}^{T},Y_i)$ identically and  independently  distributed according to distribution (\ref{eq1.0}).
Then we have the following complete likelihood  of the standard parameters according to (\ref{eq1.0})
\begin{subequations}
\begin{equation}\label{eq10_8o}
L(\{\mathbf x_{i1}^{T},\mathbf z_{i1}^T,y_i\}_{i=1}^n )=
\end{equation}
\begin{equation}\label{eq10_8b}
\prod_{t=1}^{T}\prod_{i=1}^n {\rm P}( \mathbf x_{it} \mid  \mathbf x_{i1}^{t-1}, \mathbf z_{i1}^{t-1};\Psi_{\mathbf x_{t}})
{\rm P}(z_{it}\mid  \mathbf x_{i1}^{t}, \mathbf z_{i1}^{t-1}; \Psi_{z_t})
\end{equation}
\begin{equation}\label{eq10_8a}
\prod_{i=1}^n {\rm P}(y_i\mid \mathbf{x}_{i1}^{T}, \mathbf z_{i1}^T ; \Psi_y).
\end{equation}
\end{subequations}
We will estimate the blip effect  parameter  $\boldsymbol\gamma$ based on the conditional likelihood (\ref{eq10_8a}) in this subsection and based on the complete likelihood (\ref{eq10_8o}) in the next subsection.

The distribution ${\rm P}(\mathbf{x}_1^{T}, \mathbf z_1^T)$ of   treatments and covariates is estimated by the corresponding proportion $\widehat {\rm P}(\mathbf{x}_1^{T}, \mathbf z_1^T)$. Given $\{(\mathbf x_{i1}^{T},\mathbf z_{i1}^{T})\}$, this proportion   has no variability and is denoted by ${\rm P}_c(\mathbf{x}_1^{T}, \mathbf z_1^T)$, where the subscript 'c' indicates that it is conditional on $\{(\mathbf x_{i1}^{T},\mathbf z_{i1}^{T})\}$.  Applying  ${\rm P}_c(\mathbf{x}_1^{T}, \mathbf z_1^T)$, we identify the conditional  distribution  ${\rm P}(y\mid \mathbf
x_{t},z_{t})$ by
$$
{\rm P}_c(y\mid \mathbf
x_{t},z_{t})=E_c\{{\rm P}(y\mid \mathbf x_1^{T},  \mathbf z_1^{T}   )\mid \mathbf
x_{t},z_{t}\},
$$
where the expectation is with respect to   ${\rm P}_c(\mathbf x_1^{t-1}, \mathbf z_1^{t-1},\mathbf x_{t+1}^{T},$ $ \mathbf z_{t+1}^{T}\mid \mathbf x_{t},z_t)$ obtained from ${\rm P}_c(\mathbf{x}_1^{T}, \mathbf z_1^T)$. Then,
we identify the mean
$
\mu(\mathbf
x_{t},z_{t})$ by
\begin{equation}\label{eq10.1.1}
\mu_c(\mathbf x_{t},z_t)=E_c(Y\mid \mathbf x_{t},z_t),
\end{equation}
where the expectation is with respect to  $ {\rm P}_c(y\mid \mathbf
x_{t},z_{t})$ obtained above.
According to (\ref{eq10.5}), the point effect $\theta(\mathbf
x_{t}; z_t)$ is then  identified by
$$
\theta_c(\mathbf
x_{t}; z_t)=\mu_c(\mathbf x_{t},z_t)-\mu_c(\mathbf x_{t},z_t=0).
$$
Therefore, the point effect vector
 $\boldsymbol\theta_{t}$ is identified by $\boldsymbol\theta_{c, t}$ which consists of $\theta_c(\mathbf
x_{t}; z_t)$, and $\boldsymbol\theta$ by $\boldsymbol\theta_c$ which consists of $\boldsymbol\theta_{c, t}$.
Furthermore, given the proportion ${\rm P}_c(\mathbf{x}_1^{T}, \mathbf z_1^T)$, the  $c_{j}(\mathbf x_{t}, z_t)$ and thus $\mathbf c(\mathbf x_{t}, z_t)$, $\mathbf C_t$ and $\mathbf C$  are all given  and denoted by $c_{c,j}(\mathbf x_{t}, z_t)$, $\mathbf c_c(\mathbf x_{t}, z_t)$,  $\mathbf C_{c,t}$ and $\mathbf C_c$.
According to (\ref{eq10.7}), the blip effect parameter $\boldsymbol\gamma$ is then identified by
\begin{equation}\label{eq10.9}
\boldsymbol\gamma_c=({\mathbf C}_c' {\boldsymbol \Sigma}^{-1}_c { \mathbf C}_c)^{-1} {  \mathbf C}'_c  { \Sigma}^{-1}_c {\boldsymbol\theta}_c,
\end{equation}
where $\boldsymbol \Sigma_c$ is a positive definite  matrix such that $({\mathbf C}_c' {\boldsymbol \Sigma}^{-1}_c { \mathbf C}_c)$ is invertible.

Based on the conditional likelihood (\ref{eq10_8a}), we only need to estimate ${\boldsymbol\theta}_c$ to estimate $\boldsymbol\gamma_c$.
Formula (\ref{eq10.1.1}) implies that we can estimate $\mu_c(\mathbf x_{t},z_t)$ by  taking
 the average of $y_i$ in stratum $(\mathbf
x_{t},z_{t})$. Using  $\hat \mu_c(\mathbf x_{t},z_t)$, we obtain $
\hat\theta_c(\mathbf
x_{t}; z_t)=\hat\mu_c(\mathbf
x_{t},z_t)-\hat\mu_c(\mathbf
x_{t},z_t=0)$ and then  $\widehat {\boldsymbol\theta}_{c,t}$ for all point effects at time $t$. Clearly, $\widehat {\boldsymbol\theta}_{c,t}$
is unbiased.
We may   use  standard softwares to obtain  the covariance matrix ${\rm cov}_c(\widehat{\boldsymbol\theta}_{c,t})$.

In Supplement I of  Supporting Material, we show that for the normal outcome,
\begin{equation}\label{eq10.20}
{\rm cov}_c(\widehat{\boldsymbol\theta}_{c,s}; \widehat {\boldsymbol\theta}_{c,t})=0, \qquad s \neq t,
\end{equation}
which means that conditional on $\{(
\mathbf x_{i1}^{T},\mathbf z_{i1}^{T}) \}$, the estimates of the  point effects are not correlated at different times.
For the outcome of  common distributions, the mean $\mu_c(\mathbf x_{s},  z_s)$ and thus  $\boldsymbol\theta_{c,s}$ are highly robust to  $\boldsymbol\theta_{c,t}$  at time $t > s$, so we assume  (\ref{eq10.20}) for non-normal outcomes in the following development. This assumption only leads to a minor loss of efficiency in a regression, see, for instance, Sen and Srivastava (1997). Furthermore, it is far weaker  than those assumptions
for the variance-covariance structure used in  the existing   estimation methods, for instance,  Almirall et al. (2010), Wallace et al. (2016) and Wang and Yin (2019).

Putting     $\widehat {\boldsymbol\theta}_{c,t}$ at $t=1,\dots, T$ together, we obtain the unbiased  estimate   $\widehat {\boldsymbol\theta}_{c}$ for all   point effects.
 The conditional covariance matrix  ${\rm cov}_c(\widehat {\boldsymbol\theta}_{c}) $ is then a diagonal matrix with diagonal  submatrices ${\rm cov}_c(\widehat{\boldsymbol\theta}_{c,t})$ at times $t=1,\dots, T$. Let
$\mathbf \Sigma_c=$ ${\rm cov}_c(\widehat {\boldsymbol\theta}_{c}) $.
 If $(  {\mathbf C}'_c  \mathbf \Sigma^{-1}_c {\mathbf C}_c)$ is invertible, then we
 regress  $\widehat {\boldsymbol\theta}_{c}$ on the obtained design matrix $\mathbf C_c$
according to model (\ref{eq10.6}) and
 obtain
\begin{PP}\label{P1}
Based on the conditional likelihood (\ref{eq10_8a}),  the  estimate for $\boldsymbol\gamma_c$ is
\begin{equation}\label{eq20.5}
\widehat{\boldsymbol\gamma}_c=({\mathbf C}'_c {\mathbf  \Sigma}^{-1}_c { \mathbf C}_c)^{-1} {  \mathbf C}'_c  { \mathbf \Sigma}^{-1}_c  \widehat{\boldsymbol\theta}_c.
\end{equation}
Furthermore,  the estimate is unbiased: $E_c(\widehat{\boldsymbol\gamma}_c)={\boldsymbol\gamma}_c$. The conditional covariance matrix is
\begin{equation}\label{eq20.5_1}
{\rm cov}_c(\widehat {\boldsymbol\gamma}_c)=(  {\mathbf C}'_c  \mathbf \Sigma^{-1}_c {\mathbf C}_c)^{-1}.
\end{equation}
\end{PP}

Now we study the asymptotic properties of
$$
\widehat{\boldsymbol\gamma}_{n,c}=({\mathbf C}'_{n,c} {\mathbf  \Sigma}^{-1}_{n,c} { \mathbf C}_{n,c})^{-1} {  \mathbf C}'_{n,c}  {\mathbf  \Sigma}^{-1}_{n,c} \widehat{\boldsymbol\theta}_{n,c},
$$
where $n$ indicates the sample size.
Noticeably,   ${\boldsymbol\theta}_{n,c,t}$ at time $t=1,\dots, T$  is an estimand in single-point causal inference.  The conditions for the consistency and asymptotic normality  of its  estimate $\widehat{\boldsymbol\theta}_{n,c,t}$ are well studied in single-point causal inference and satisfied in most practices, see, for instance, Fahrmeir  and Tutz (1994).

If  $\widehat{\boldsymbol\theta}_{n,c,t}$ is consistent,  so is
$\widehat{\boldsymbol\theta}_{n,c}$. Therefore  $\widehat{\boldsymbol\gamma}_{n,c}$
is  consistent with ${\boldsymbol\gamma}_{n,c}$ given by (\ref{eq10.9}).
If the estimate  $\widehat {\boldsymbol\theta}_{n,c,t}$ is   asymptotically normal:
$\sqrt{n}(\widehat{\boldsymbol\theta}_{n,c,t}  -\boldsymbol\theta_{n,c,t})
\overset{d}{\longrightarrow}  N\{\mathbf 0, n{\rm cov  }_c(\widehat{\boldsymbol\theta}_{n,c,t}) \}$,  so is $\widehat{\boldsymbol\theta}_{n,c}$:
$\sqrt{n}(\widehat{\boldsymbol\theta}_{n,c}  -\boldsymbol\theta_{n,c})\overset{d}{\longrightarrow}  N\{\mathbf 0, n{\rm cov  }_c(\widehat{\boldsymbol\theta}_{n,c}) \}$.
 Therefore $\widehat {\boldsymbol\gamma}_{n,c}$ is also asymptotically normal:
$$
\sqrt{n}(\widehat{\boldsymbol\gamma}_{n,c}-\boldsymbol\gamma_{n,c}) \overset{d}{\longrightarrow} N\{\mathbf 0, n{\rm cov}_c(\widehat {\boldsymbol\gamma}_{n,c})\}.
$$
However, even if $\boldsymbol\gamma_{n,c}$ may converge to
$\boldsymbol\gamma$, the asymptotic normal distribution of $\widehat{\boldsymbol\gamma}_{n,c}$  cannot be used to construct the Wald test for  hypotheses on $\boldsymbol\gamma$, because this distribution does not incorporate the variability of   treatments and covariates.  However,  in the next subsection, we will use
 $\widehat{\boldsymbol\gamma}_{n,c}$ and ${\rm cov}_c(\widehat {\boldsymbol\gamma}_{n,c})$ to obtain the marginal estimate of
 $\boldsymbol\gamma$  and its covariance matrix based on the complete likelihood (\ref{eq10_8o}) incorporating the variability of    treatments and covariates.

\subsection{Marginal estimate of blip effect parameter}
Based on the complete likelihood (\ref{eq10_8o}), the  estimates of
$\mu(\mathbf x_{t},z_t)$, $\theta(\mathbf
x_{t}; z_t)$, ${\boldsymbol\theta}_t$, $ {\boldsymbol\theta}$, $ c_j(\mathbf x_{t}, z_t)$, $\mathbf c(\mathbf x_{t}, z_t)$, $\mathbf C_t$, $\mathbf C$, and finally $ {\boldsymbol\gamma}$
are marginal estimates   and  denoted by  $\hat\mu(\mathbf x_{t},z_t)$, $\hat\theta(\mathbf
x_{t}; z_t)$, $\widehat {\boldsymbol\theta}_t$, $\widehat {\boldsymbol\theta}$, $\hat c_j(\mathbf x_{t}, z_t)$, $\widehat \mathbf c(\mathbf x_{t}, z_t)$, $\widehat \mathbf C_t$, $\widehat \mathbf C$, and finally $\widehat {\boldsymbol\gamma}$.

Formula (\ref{eq10.3.1}) implies that we can obtain the  estimate $\hat \mu(\mathbf x_{t},z_t)$ by taking   the average of $y_i$ in stratum $(\mathbf
x_{t},z_{t})$.
Using  $\hat \mu(\mathbf x_{t},z_t)$, we obtain  $\hat\theta(\mathbf
x_{t}; z_t)= \hat \mu(\mathbf x_{t},z_t)- \hat \mu(\mathbf x_{t},z_t=0)$ and then  $\widehat{\boldsymbol\theta}_t$ and  $\widehat{\boldsymbol\theta}$.
Comparing with $\hat\mu_c(\mathbf
x_{t},z_t)$ obtained from the previous subsection, we see $\hat\mu(\mathbf
x_{t},z_t)=\hat\mu_c(\mathbf
x_{t},z_t)$, both being the average of $y_i$ in stratum $(\mathbf
x_{t},z_t)$.  Therefore,  we have $\hat\theta(\mathbf
x_{t}; z_t)$ $=\hat\theta_c(\mathbf
x_{t}; z_t)$, and then $\widehat{\boldsymbol\theta}_t=\widehat{\boldsymbol\theta}_{c,t}$ and $\widehat{\boldsymbol\theta} =\widehat{\boldsymbol\theta}_c$.

Given the proportion   $\widehat {\rm P}(\mathbf{x}_1^{T}, \mathbf z_1^T)$, then the estimates  $\hat c_j(\mathbf x_{t}, z_t)$, $\widehat \mathbf c(\mathbf x_{t}, z_t)$, $\widehat \mathbf C_t$, $\widehat \mathbf C$ are given; here we use  ' $\hat{}$ ' to emphasize that these estimates possess variabilities of   treatments and covariates, namely, they have non-zero variances.
Because $\widehat {\rm P}(\mathbf{x}_1^{T}, \mathbf z_1^T)$ $={\rm P}_c(\mathbf{x}_1^{T}, \mathbf z_1^T)$,
 we have
  $\hat c_j(\mathbf x_{t}, z_t)$ $= c_{c,j}(\mathbf x_{t}, z_t)$. Therefore,  we have  $\widehat {\mathbf c}(\mathbf x_{t}, z_t) = {\mathbf c}_c(\mathbf x_{t}, z_t)$, $\widehat {\mathbf C}_t=\mathbf C_{c,t}$  and $\widehat {\mathbf C}=\mathbf C_{c}$.

Let $\mathbf \Sigma_t=E\{{\rm cov  }_c(\widehat{\boldsymbol\theta}_{c,t})\}$  with respect to the probability ${\rm P}
(\mathbf x_{1}^{T},\mathbf z_{1}^{T})$. Then an estimate of ${\mathbf \Sigma}_t$ is ${\rm cov  }_c(\widehat{\boldsymbol\theta}_{c,t})$: $\widehat {\mathbf \Sigma}_t={\rm cov  }_c(\widehat{\boldsymbol\theta}_{c,t})$.
Take
$\mathbf \Sigma$ as a diagonal matrix with diagonal submatrices $\mathbf \Sigma_t$ at times $t=1,\dots, T$.
Recalling that $\mathbf \Sigma_c$  is the diagonal matrix with diagonal  submatrices ${\rm cov}_c(\widehat{\boldsymbol\theta}_{c,t})$ at times $t=1,\dots, T$, we have that an estimate of
${\mathbf \Sigma}$ is
 $\mathbf \Sigma_{c}$:
$\widehat \mathbf \Sigma =\mathbf \Sigma_{c}$.

Now from (\ref{eq10.7}) and (\ref{eq20.5}),   we see $\widehat{\boldsymbol\gamma}=\widehat{\boldsymbol\gamma}_c$, implying that the marginal estimate of $\boldsymbol\gamma$ is equal to the conditional estimate of $\boldsymbol\gamma_c$. However, the marginal covariance matrix ${\rm cov}(\widehat{\boldsymbol\gamma})$ which incorporates the variability of $\{(
\mathbf X_{i1}^{T},\mathbf Z_{i1}^{T}) \}$ is not equal to the conditional covariance matrix ${\rm cov}_c(\widehat{\boldsymbol\gamma}_c)$ which is condition on $\{(
\mathbf X_{i1}^{T},\mathbf Z_{i1}^{T}) \}$ and does not have such a variability.

Summarizing the above observations and applying
 Proposition \ref{P1}, we obtain
\begin{PP}\label{P2}
Based on the complete likelihood (\ref{eq10_8o}),
the marginal estimate of $\boldsymbol\gamma$     is
\begin{equation}\label{eq20.6}
\widehat{\boldsymbol\gamma}=(\widehat {\mathbf C}' \widehat {\boldsymbol \Sigma}^{-1} \widehat {\mathbf C})^{-1} \widehat {\mathbf C}' \widehat {\boldsymbol\Sigma}^{-1}  \widehat{\boldsymbol\theta},
\end{equation}
where $\widehat{\boldsymbol\theta}=\widehat{\boldsymbol\theta}_c$,  $\widehat {\mathbf C}={\mathbf C}_c$, and $\widehat {\boldsymbol \Sigma}=\boldsymbol\Sigma_c$, which are based on  the conditional likelihood (\ref{eq10_8a}) and given in the previous subsection. However, the marginal estimate $\widehat {\boldsymbol\gamma}$ is biased with $\boldsymbol\gamma$:
$E(\widehat {\boldsymbol\gamma})\ne \boldsymbol\gamma $. The marginal covariance matrix of $\widehat {\boldsymbol\gamma}$ is  equal to
$$
{\rm cov}(\widehat{\boldsymbol\gamma})=E\{ {\rm cov}_c(\widehat{\boldsymbol\gamma}_c)  \} + {\rm cov}(\boldsymbol\gamma_c),
$$
in which ${\rm cov}_c(\widehat{\boldsymbol\gamma}_c)$ is given by  (\ref{eq20.5_1})
and $\boldsymbol\gamma_c$ by (\ref{eq10.9}).
\end{PP}
\noindent {\bf Proof:}   The bias is due to the Jensen's inequality:
$$
E\{(\widehat {\mathbf C}' \widehat {\boldsymbol \Sigma}^{-1} \widehat {\mathbf C})^{-1} \widehat {\mathbf C}' \widehat {\boldsymbol\Sigma}^{-1}  \widehat{\boldsymbol\theta} \}
$$
$$
\neq
[ \{E(\widehat {\mathbf C})\}' \{E(\widehat {\boldsymbol \Sigma})\}^{-1} E(\widehat {\mathbf C}) ]^{-1} \{E(\widehat {\mathbf C})\}' \{E(\widehat {\boldsymbol\Sigma})\}^{-1}E(\widehat{\boldsymbol\theta}).
$$
By the law of total covariance,  we can decompose the marginal covariance matrix ${\rm cov}(\widehat{\boldsymbol\gamma})$  into  two terms. The first term is  the mean  of the  conditional covariance matrix $ {\rm cov}_c(\widehat{\boldsymbol\gamma}) $. Due to $\widehat {\boldsymbol\gamma}=\widehat {\boldsymbol\gamma}_c$, we have
$ {\rm cov}_c(\widehat{\boldsymbol\gamma}) $  $= {\rm cov}_c(\widehat{\boldsymbol\gamma}_c) $ given by (\ref{eq20.5_1}). Thus the first term is
$E\{ {\rm cov}_c(\widehat{\boldsymbol\gamma}_c)  \}$.
The second term is
 the covariance matrix of the conditional mean of $\widehat {\boldsymbol\gamma}$.  Due to $\widehat {\boldsymbol\gamma}=\widehat {\boldsymbol\gamma}_c$,
the conditional mean of $\widehat {\boldsymbol\gamma}$
  is equal to the conditional mean of $\widehat {\boldsymbol\gamma}_c$, which is equal to $ {\boldsymbol\gamma}_c$
  given by (\ref{eq10.9}). Thus, the second term is ${\rm cov}(\boldsymbol\gamma_c)$.
 Putting the two terms together, we obtain the formula for ${\rm cov}(\widehat{\boldsymbol\gamma})$ in the proposition.

The bias of $\widehat {\boldsymbol\gamma}$  is   small and  can be ignored in  practical situations, where $\mathbf C$ usually varies slowly with $(\mathbf{x}_1^{T}, \mathbf z_1^T)$.  The small bias will also be illustrated by simulation in Section $4$. To estimate  the marginal covariance matrix ${\rm cov}(\widehat{\boldsymbol\gamma})$ in practice,
 we can apply the bootstrap method to formula (\ref{eq20.6}).

Now we study the asymptotic properties of
$$
\widehat{\boldsymbol\gamma}_n=(\widehat {\mathbf C}'_n \widehat {\boldsymbol \Sigma}^{-1}_n \widehat {\mathbf C})^{-1}_n \widehat {\mathbf C}'_n \widehat {\boldsymbol\Sigma}^{-1}_n  \widehat{\boldsymbol\theta}_n,
$$
where $n$ indicates the sample size.
Noticeably,   ${\boldsymbol\theta}_{n,t}$,  ${\mathbf \Sigma}_{n,t}$ and ${\mathbf C}_{n,t}$ at time $t=1,\dots, T$  are  the estimands in single-point causal inference.  The conditions for the consistency and asymptotic normality  of their estimates are well studied in single-point causal inference and satisfied in most practices, see, for instance, Fahrmeir  and Tutz (1994).

If $\widehat {\boldsymbol{\theta}}_{n,t}$ is consistent at each time $t=1,\ldots, T$, so is $\widehat {\boldsymbol{\theta}}_{n}$.
If  $n\widehat {\mathbf \Sigma}_{n,t}$ is consistent, so is   $n\widehat {\mathbf \Sigma}_{n}$. If $\widehat {\mathbf C}_{n,t}$ is consistent, so is $\widehat {\mathbf C}_{n}$. Therefore $\widehat {\boldsymbol\gamma}_n$ is consistent with
$$
\boldsymbol\gamma=({\mathbf C}' { \mathbf \Sigma}^{-1} { \mathbf C})^{-1} {  \mathbf C}'  { \mathbf \Sigma}^{-1} \boldsymbol{\theta}.
$$
Furthermore, if $\widehat {\mathbf C}_{n,t}$ is asymptotically normal at each time $t=1,\ldots, T$:
$$\sqrt{n}(\widehat {\mathbf C}_{n,t}-\mathbf C_t) \overset{d}{\longrightarrow}
N[\mathbf 0, n{\rm cov}\{\widehat {\mathbf C}_{n,t}\}],
$$
then  $\widehat {\mathbf C}_{n}$ is asymptotically normal:
$$\sqrt{n}(\widehat {\mathbf C}_{n} -\mathbf C )  \overset{d}{\longrightarrow}
N[\mathbf 0, n{\rm cov}\{\widehat {\mathbf C}_{n} \}].
$$
 If $\widehat {\boldsymbol\theta}_{n,t}$ is asymptotically normal:
$$
\sqrt{n}(\widehat {\boldsymbol\theta}_{n,t}  - {\boldsymbol \theta}_t) \overset{d}{\longrightarrow}
N[\mathbf 0, n{\rm cov}\{\widehat {\boldsymbol \theta}_{n,t}\}],
$$
 then $\widehat {\boldsymbol \theta}_{n}$ is also asymptotically normal:
$$
\sqrt{n}(\widehat {\boldsymbol\theta}_{n} - {\boldsymbol\theta}) \overset{d}{\longrightarrow}
N[\mathbf 0, n{\rm cov}\{\widehat {\boldsymbol \theta}_{n} \}].
$$
Because $\widehat{\boldsymbol\gamma}_n$ is a smooth function of $\widehat \mathbf C_n$, $\widehat {\boldsymbol\Sigma}_n$ and $\widehat {\boldsymbol\theta}_n$, also noticing that  $n\widehat {\boldsymbol\Sigma}_n$ and $\{\widehat{\mathbf C}_n' (n\widehat{\boldsymbol \Sigma}_n)^{-1} \widehat{ \mathbf C}_n\}$  are invertible by study design,  we have that $\widehat {\boldsymbol\gamma}_n$ is asymptotically normal:
  \begin{equation}\label{eq30.1}
\sqrt{n}(\widehat{\boldsymbol\gamma}_n -\boldsymbol\gamma)\overset{d}{\longrightarrow} N\{\mathbf 0, n{\rm cov}(\widehat {\boldsymbol\gamma}_n)\}.
\end{equation}

Noticeably, the assumptions  above on $\widehat {\boldsymbol\theta}_{n,t}$,  $\widehat {\mathbf \Sigma}_{n,t}$ and $\widehat {\mathbf C}_{n,t}$
 may be   weakened  for the consistency and asymptotic normality of $\widehat {\boldsymbol \gamma}$, because
 there are
far more  point effects  from   a treatment sequence than from a single-point treatment.   More point effects imply more information about $\boldsymbol \gamma$. Although it is of considerable interest particularly for a long treatment sequence, the issue is  beyond the scope of this article and will not be further investigated here.

\subsection{Wald test}
Recall the  hypothesis  (\ref{eq1.5}), that is,
$$
H_0:  \mathbf H\boldsymbol\gamma - \boldsymbol\rho=\mathbf 0  \quad {\mbox {\rm against}} \quad H_1:   \mathbf H\boldsymbol\gamma -\boldsymbol\rho \neq \mathbf 0,
$$
where  $\mathbf H$ is a $l\times k$ matrix with $l \leq k$ and $\boldsymbol\rho$ is a constant $l$-dimensional vector.
Applying  the marginal   estimate $\widehat {\boldsymbol\gamma}$ and its covariance matrix ${\rm cov}(\widehat {\boldsymbol\gamma})$  given by Proposition \ref{P2} in the previous subsection,
we obtain the Wald statistic for the hypothesis   as
\begin{equation}\label{eq30.2}
W=(\mathbf H\widehat {\boldsymbol\gamma}-\boldsymbol\rho)' \{\mathbf H{\rm cov}(\widehat {\boldsymbol\gamma})\mathbf H'\}^{-1}
(\mathbf H\widehat {\boldsymbol\gamma}-\boldsymbol\rho).
\end{equation}
\begin{T}\label{T1}
Suppose that  $\widehat{\boldsymbol\gamma}$ is asymptotically normal, namely,
formula  (\ref{eq30.1}) is true. Then under  the null  hypothesis $H_0$, the Wald statistic $W$    has a limiting   $\chi^2_l$ distribution with $l$ degrees of freedom. Under the alternative hypothesis $H_1$, the $W$ has a   limiting noncentral  $\chi^2_{l, \lambda}$ distribution with $l$ degrees of freedom and the noncentrality  parameter $\lambda$ arising from $\mathbf H\boldsymbol\gamma -\boldsymbol\rho \neq \mathbf 0$.
\end{T}
For a given significance level $\alpha$, the null hypothesis is rejected  if $W$ exceeds the upper $100(1-\alpha)\%$ quantile of the $\chi^2_{l}$ distribution.

The obtained Wald test  has  the following advantages over the likelihood ratio test and the score test.
\begin{R}\label{R1}
First,  the blip effect parameter  $\boldsymbol\gamma$ is estimated via a small number of point effects $\theta(\mathbf x_{t}; z_t  )$, so the curse of dimensionality does not necessarily occur. Second,  model (\ref{eq10.6}) is an unsaturated model  for these point effects, so the null paradox does not necessarily occur. Third,   model (\ref{eq10.6}) allows for estimating $\boldsymbol\gamma$ under the null hypothesis,  so a high-dimensional constraint on standard parameters has been avoided.
\end{R}

\subsection{Procedure of conducting Wald test}
In practice, we can conduct the Wald test in the following four stages. In the first stage, we  find the treatment assignment condition and conduct  initial assessment of the point effects and SNMM. In the second stage, we decompose the point effects into the blip effects to obtain a model for the point effects, as described in Section $3.1$. In the third stage, we apply the model to estimate the blip effect parameter, as described in  Section $3.2$.
In the fourth  stage, we apply the bootstrap method to the third stage to estimate the marginal covariance matrix for the blip effect parameter, as described in Section $3.3$,
  calculate the Wald statistic and conduct the hypothesis test as described in Section $3.4$.
 This procedure will be illustrated by a  simulation study in Section $4$ and a real medical study in Section $5$.

\section{Simulation study}
The treatment sequence  has a length of $T=3$. The treatment variables are dichotomous with  $Z_t=0,1$ ($t=1,2,3$). For simplicity, we do not include  the stationary covariate $X_1$ in the simulation. The  time-dependent covariates are polytomous   with $X_{t}=0, 1,2,3$ ($t=2,3$).  After the last treatment $Z_3$, there is  an outcome variable $Y$ of interest. A summary of the variables is $(Z_1,X_2, Z_2, X_3, Z_3, Y)$ in the temporal order, with their  realizations   $(z_1,x_2, z_2, x_3, z_3, y)$. Conditional on $(z_1,x_2, z_2, x_3, z_3)$, the outcome $Y$
follows the normal,  Bernoulli or Poisson distribution.

Suppose SNMM of  the following form. At $t=1$, there is only one blip effect of $z_1=1$:  $\phi(z_1=1)   =\gamma_1$. At $t=2$, there are  four blip effects  of $z_2=1$ depending only on $x_2=0,1,2,3$:
$\phi(z_1,   x_2=j;z_2=1  )=\gamma_{2j}$,   $j=0,1,2,3$. At $t=3$, there are  four blip effects  of $z_3=1$ depending only on $ x_3=0,1,2,3$:
$\phi(z_1,   x_2, z_2, x_3=j; z_3=1  )=\gamma_{3j}$,  $j=0,1,2,3$. Then for this SNMM, we have the blip effect parameter $\boldsymbol\gamma=(\gamma_1, \gamma_{20},\gamma_{21}, \gamma_{22}, \gamma_{23},
\gamma_{30}, \gamma_{31}, \gamma_{32}, \gamma_{33})'$. Denote the true value of $\boldsymbol\gamma$ by $\boldsymbol\gamma_0=
(\gamma_{1,0}, \gamma_{20,0},\gamma_{21,0}, \gamma_{22,0}, \gamma_{23,0},
\gamma_{30,0}, \gamma_{31,0}, \gamma_{32,0}, \gamma_{33,0})'$.
 For   normal outcome,  we set the true value  $\boldsymbol\gamma_0=(2, 3,-4, -4,3,3,-4,-4, 3)'$;
for dichotomous outcome, $\boldsymbol\gamma_0= (-0.2, 0.1, -0.15, -0.15,$ $ 0.1, 0.1, -0.15, -0.15, 0.1)'$;
for Poisson outcome, $\boldsymbol\gamma_0=(2, 4, -3,  -3, 4, 4,-3, -3, 4)'$.

The treatment assignment satisfies (\ref{eq10.4}),   that is, the assignment of $z_2$   depends only on  $x_2$ and that of $z_3$ only on $ x_3$.
Then,
we have a total of nine point   effects: one  $
\theta(z_1=1)$ of $z_1=1$ at time $t=1$,   four $
\theta(
x_2; z_2=1)$ of $z_2=1$ with $
x_2=0,1,2,3$ at $t=2$, and four $\theta(
x_3; z_3=1)$ of $z_3=1$ with $\
x_3=0,1,2,3$ at $t=3$.
In Supplement II of  Supporting  Material, we construct
the data-generating mechanism that corresponds to the treatment assignment condition and the blip effect parameter above.

We  test $10$ hypotheses labeled by $A$ through $J$. The null hypothesis $A_0$:  $\gamma_1=\gamma_{1,0}$; the first  alternative
$A_1$:  $\gamma_1=\gamma_{1,0}+c$; the second alternative $A_2$:  $\gamma_1=\gamma_{1,0}+2c$; where $c=1$ for the normal and Poisson outcomes and $c=0.1$ for the dichotomous outcome.
$B_0$ through $E_0$:  $\gamma_{2j}=\gamma_{2j,0}$, $j=0,1,2,3$, respectively;
$B_1$ through $E_1$:  $\gamma_{2j}=\gamma_{2j,0}+c$;  $B_2$ through $E_2$:  $\gamma_{2j}=\gamma_{2j,0}+2c$.
$F_0$ through $I_0$:  $\gamma_{3j}=\gamma_{3j,0}$, $j=0,1,2,3$,  respectively;
$F_1$ through $I_1$:  $\gamma_{3j}=\gamma_{3j,0}+c$;  $F_2$ through $I_2$:  $\gamma_{3j}=\gamma_{3j,0}+2c$. In particular, we also test the equalities between the blip effects at times $t=2, 3$, that is,
$J_0$:
$\gamma_{2j}=\gamma_{3j}$, $j=0,1,2,3$; $J_1$:
$\gamma_{2j}=\gamma_{3j}+c$; $J_2$:
$\gamma_{2j}=\gamma_{3j}+2c$.

The sample sizes are chosen as $1000$, $2000$ and $3000$, such that the point effects are estimable. For every sample size,
 $1000$  data sets are generated to simulate the type I and II errors of each outcome type.
The  covariance matrix for  the   blip effect parameter is estimated  by  using the basic bootstrap method with only $500$ replications  due to our limited computing power.
In Supplement II of  Supporting  Material, we describe the simulation study in detail.
The relevant SAS codes used for the simulation are given in Data and Codes of the article.

Table $1$  presents the type I and II error rates of hypothesis $A$ through $J$ at the significance level of $0.05$. From rate0 of the null hypothesis in the table, we see that the Wald test nearly achieves the nominal level  of  type I error, despite a crude bootstrap method for  the covariance matrix of the blip effect parameter.   From rate1 of the first alternative hypothesis and rate2 of the second alternative hypothesis, we  see that the type II error  decreases with an increasing  sample size and an increasing   difference between  the alternative and null hypotheses. Comparing columns  $A$--$E$ with $F$--$I$, we find that the Wald test possesses  no less powers  for the blip effects of the earlier  treatments ($z_1=1$ and $z_2=1$) than the later treatment ($z_3=1$). Considering  the small difference between the null and alternative hypotheses ($1$ or $2$ for the normal and Poisson outcomes and $0.1$ or $0.2$ for the dichotomous outcome), the Wald test is    powerful in testing the blip effects.

Table $2$ presents the estimate and its variance  for the blip effect parameter obtained under no constraint or under the null hypothesis $J_0$. For the sake of space, we use only one sample size of  $1000$. From this table, we see that the bias of the estimated blip effect parameter is negligible.  We also see   a considerable reduction in variance under the null hypothesis $J_0$ compared to  no constraint.

\section{Medical example}
In Sweden, patients  usually seek medical help at  hospitals near their residential areas. When cancer is diagnosed,   they may   stay at the diagnosing hospital or transfer to another hospital for treatment. The hospital diagnosing the cancer is called home hospital  while the one treating the cancer is called treating hospital.
The  performance of the home and treating hospitals is of considerable interest  to patients,  doctors and   public health agencies.

Here, we study which types of the home and treating  hospitals, large versus small, perform better on cancer survival, where the type is determined by the number of patients received there. The data contains the information of $1070$ stomach cancer patients  from a clinical study  during a period between $1988$ and $1995$ in hospitals
located in central and northern Sweden. Stomach  cancer is highly malignant with bad prognosis and
its one-year survival is a good measure of the performance of both home and treating hospital types.

The home hospital is the treatment variable $Z_1$ at time $t=1$: $z_1=1$ for large type and $z_1=0$ for small type. The treating hospital is $Z_2$ at $t=2$:  $z_2=1$ for large type and $z_2=0$ for small type. The outcome of a patient is $Y$: $y=1$ for a successful one-year survival and $y=0$ otherwise. The following stationary covariates before $Z_1$ were measured:
gender ($X_{11}$),  geographic area ($X_{12}$) and age ($X_{13}$).
 Gender was $x_{11}
= 1$ for male and $x_{11}
= 0$ for female. Geographic area was categorized into urban ($x_{12}
= 1$) versus rural
($x_{12}
= 0$).
 Age was a
continuous variable.
The time-dependent covariate between $Z_1$ and $Z_2$ was
cancer stage ($X_2$), which
 was
 categorized into  the advanced stage ($x_2
= 1$) and the early stage ($x_2
= 0$). The  data   is given in
 Data and Codes of the article together with the SAS code for the analysis.
  In the following, we will test if the blip effects of $z_1=1$ and $z_2=1$ are   zeros respectively.

Due to a long-term social welfare system and relatively uniform culture in the country,  the assumption of no unmeasured confounders is approximately  true for home hospital  $Z_1$,
 at least after conditioning on   $(x_{11}, x_{12}, x_{13})$ according to the medical experts. Similarly, the assumption is also approximately  true for treating hospital $Z_2$ conditional  on
$(x_{11}, x_{12}, x_{13}, z_1,x_2)$.

In the first stage, we   conduct  initial assessment of the point effects and  SNMM  by modeling the means $\mu( x_{11},  x_{12}, x_{13},z_{1})$ and $\mu( x_{11},  x_{12}, x_{13},z_{1},  x_2, z_2)$ in combination with the subject matter knowledge. It is difficult  to specify the distribution
${\rm P}(y\mid  x_{11},  x_{12},x_{13}, z_{1})$ due to influences of $X_2$ and $Z_2$, so we use the usual quasi likelihood approach to modeling $\mu( x_{11},  x_{12},x_{13}, z_{1})$, see, for instance, Fahrmeir  and Tutz (1994).
The distribution ${\rm P}(y\mid  x_{11},  x_{12},x_{13}, z_{1},  x_2, z_2)$ is binomial, which  is used to model
$\mu( x_{11},  x_{12}, x_{13},z_{1},  x_2, z_2)$.
By the usual  significance test  at the significance level of $0.1$,   we model $\mu( x_{11}, x_{12}, x_{13}, z_{1})$,  exclude the non-significant variables $X_{12}$, and  obtain
\begin{equation}\label{eq100}
\mu( x_{11}, x_{13},  z_{1})=\beta_0+x_{11}\beta_{1}+x_{13}\beta_{2}+  z_{1}\theta_1,
\end{equation}
where $\theta_1=\theta(x_{11}, x_{13} ; z_1=1)$ is   the point  effect     of $z_1=1$, which is the same for all $(x_{11},x_{13})$.
We also model $\mu( x_{11},   x_{12},x_{13},  z_{1}, x_2, z_{2})$,  exclude the non-significant variables  $X_{12}$, $X_{13}$, and obtain
\begin{equation}\label{eq105}
\left\{
\begin{array}{l}  \mu( x_{11}, z_1, x_2=0,  z_2)=\beta_3+
x_{11}\beta_4 + z_1\beta_5 +z_{2}\theta_{20} ,  \\
\mu( x_{11},z_1,  x_2=1, z_2)=\beta_6+
x_{11}\beta_7 +z_1\beta_8  +z_{2}\theta_{21} ,
\end{array}
\right.
\end{equation}
where $\theta_{20}=\theta(x_{11},z_1, x_2=0; z_2=1)$ is   the point  effect of $z_2=1$ when $x_2=0$ and   $\theta_{21}=\theta(x_{11},z_1, x_2=1; z_2=1)$ is the point  effect of $z_2=1$ when $x_2=1$, which  are the same for all $( x_{11},z_1)$.
Combining  with the subject knowledge  from medical experts, we set an initial SNMM as
$$
\left\{
\begin{array}{l} \phi(x_{11}, x_{13}; z_{1}=1)=\gamma_1 ,  \\
\phi(x_{11},x_{13},  z_1,  x_2=0; z_2=1)=\gamma_{20}, \\
\phi(x_{11}, x_{13}, z_1, x_2=1; z_2=1)=\gamma_{21}.
\end{array}
\right.
$$
Here, we have the blip effect parameter $\boldsymbol \gamma=(\gamma_1, \gamma_{20}, \gamma_{21})'$.

In the second stage,  we decompose the point effects into the blip effects. The point effect  $\theta_1$ of $z_1=1$ is a sum of contributions from  home hospital $z_1=1$ and    treating hospital  $z_2=1$. Because $Z_2$ is the last treatment variable, the point effect of $z_2=1$ is equal to the blip effect of $z_2=1$, that is, $\theta_{20}=\gamma_{20}$ and $\theta_{21}=\gamma_{21}$. As a result, we have
\begin{equation}\label{eq115}
\left\{ \begin{array}{l}  \theta_1= \gamma_1 +\gamma_{20}c_{20} + \gamma_{21}c_{21}, \\
 \theta_{20}=\gamma_{20}, \\
  \theta_{21}=\gamma_{21},
\end{array} \right.
\end{equation}
where $c_{20}={\rm P}(x_2=0, z_2=1\mid z_1=1)- {\rm P}(x_2=0, z_2=1\mid z_1=0)$ and $c_{21}={\rm P}(x_2=1, z_2=1\mid z_1=1)- {\rm P}(x_2=1, z_2=1\mid z_1=0)$.

In the third stage, we estimate the   blip effect parameter  $\boldsymbol \gamma=(\gamma_1, \gamma_{20}, \gamma_{21})'$. The parameters $\theta_{c,1}$, $\theta_{c,20}$ and $\theta_{c,21}$ are estimated by applying  models (\ref{eq100}) and (\ref{eq105}), where as in the earlier sections,  the subscript $c$ is added to all parameters indicating that they are identified conditional on   treatments and covariates. The estimation is based on
the binomial distribution ${\rm P}(y\mid  x_{11},  z_{1},  x_2, z_2)$ for both (\ref{eq100}) and (\ref{eq105}) and is implemented by keeping the dispersion parameter unchanged, that is,  equal to  one.

The probabilities
${\rm P}(x_2, z_2\mid z_1)$ are estimated by the corresponding proportions $\widehat {\rm P}(x_2, z_2\mid z_1)$, so we obtain $c_{c,20}$ and $c_{c,21}$.  By regression $\hat\theta_{c,1}$, $\hat\theta_{c,20}$ and $\hat\theta_{c,21}$ on one,  $c_{c,20}$ and $c_{c,21}$ according to  (\ref{eq115}), we obtain the estimates  $\hat\gamma_{c,1}$, $\hat\gamma_{c,20}$ and $ \hat \gamma_{c,21}$. According to Proposition \ref{P2},  these estimates  are respectively  equal to $\hat\gamma_{1}$, $\hat\gamma_{20}$ and $\hat \gamma_{21}$, which are based on the complete likelihood of all treatments, covariates and outcomes.

In the fourth stage, we apply the bootstrap method to the third stage above to obtain the marginal covariance matrix of  $\hat\gamma_1$, $\hat\gamma_{20}$ and $\hat \gamma_{21}$.
Then we apply (\ref{eq30.2}) to calculate the Wald statistics for three hypotheses, (1) $\gamma_1=0$ against $\gamma_1\neq 0$, (2)  $\gamma_{20}=0$ against $\gamma_{20}\neq 0$, and (3)   $\gamma_{21}=0$ against $\gamma_{21}\neq 0$. Finally we apply
 Theorem \ref{T1} to test these hypotheses.

In Table $3$, we present
the estimates of these blip effects as well as the  $95$\% CI and the p-values  for testing the three hypotheses. For comparison, we also present the results about the point effects.
Most interestingly,  from the blip effect  $\gamma_{1}$,  we see that the small home hospital might perform better than the large home hospital, indicating that the early diagnosis of stomach cancer depends more on the short waiting queue  and awareness typically at small home hospitals than on the advanced technology at large home hospitals. However, from the point effect  $\theta_{1}$, we see that the large and small home hospital might perform equally well, and this is misleading, because the point effect is a sum of contributions from the home and the   treating hospitals.

\section{Conclusion}
Due to the great need in  medical and economic researches, sequential causal inference is one of the most active areas in statistics (Hernan and Robins, 2018; An and Ding, 2018). In recent years, considerable progress has been made  in developing various estimation methods  (Hernan and Robins, 2018; An and Ding, 2018). Despite a huge volume of the  literature on methodology, however, there are few applications of sequential causal inference    in real medical and economic researches. In sharp contrast,
single-point causal inference plays a central role in many of these researches.

One possible  reason is that some statistical tools are yet to be developed, for instance, the hypothesis testing  of the blip effects, which helps to find  a pattern for the blip effects (namely, the structural nested mean model).
In this article, we found that the hypothesis test on the blip effects can be conducted via the point effects of  treatments in the sequence. Using the fact that  the point effect is simply the point effect of  treatment in single-point causal inference, we have extended the Wald test from single-point causal inference to sequential causal inference. Our method does not need  more assumptions than single-point causal inference and therefore should have a broad applicability.  It is also easy to implement in practice, as illustrated in the medical example of this article.

Due to the scope of this article, we only considered the hypothesis testing for three basic outcome types, the normal, dichotomous and Poisson  outcomes. We also restricted to additive point effects and additive blip effects. On the other hand, methods are available for testing non-additive point effects of the outcome of various types in single-point causal inference.  We believe that  our testing method can be extended  to more complex settings in the context of a treatment sequence.

\section*{Supporting  Material}
The material contains  two supplements: Supplement I for  proofs of formulas (\ref{eq10.5.1}), (\ref{eq10.6.0}) and (\ref{eq10.20}) and
Supplement  II for a description of the simulation study in Section $4$.

\pagebreak

\begin{center}
Supporting material  to "Hypothesis Testing of Blip Effects in Sequential Causal Inference"
\end{center}
\begin{center}
Xiaoqin Wang and Li Yin
\end{center}

\section*{Supplement I: proofs of  (\ref{eq10.5.1}), (\ref{eq10.6.0}) and (\ref{eq10.20})}
{\bf Proof of formula  (\ref{eq10.5.1})}:
Please notice that formula  (\ref{eq10.4}) implies
$$
{\rm P}(\mathbf x_1^{t-1}, \mathbf z_1^{t-1}\mid \mathbf x_{t}, z_t)={\rm P}(\mathbf x_1^{t-1}, \mathbf z_1^{t-1}\mid \mathbf x_{t}, z_t=0).
$$
Averaging  (\ref{eq10.0}), that is,
$$\theta(\mathbf{x}_{1}^{t}, \mathbf{z}_{1}^{t-1}; z_t)=\mu(\mathbf x_1^{t}, \mathbf z_1^{t-1},z_t)- \mu(\mathbf x_1^{t}, \mathbf z_1^{t-1},z_t=0),$$
with respect to ${\rm P}(\mathbf x_1^{t-1}, \mathbf z_1^{t-1}\mid \mathbf x_{t}, z_t)$,
we obtain
$$E\{\theta(\mathbf{x}_{1}^{t}, \mathbf{z}_{1}^{t-1}; z_t)\mid \mathbf
x_{t}, z_t\}=
$$
$$E\{\mu(\mathbf x_1^{t}, \mathbf z_1^{t-1},z_t)\mid \mathbf
x_{t}, z_t\}-E\{\mu(\mathbf x_1^{t}, \mathbf z_1^{t-1},z_t=0)\mid \mathbf
x_{t}, z_t=0\},
$$
where the last expectation is noticeably with respect to ${\rm P}(\mathbf x_1^{t-1}, \mathbf z_1^{t-1}\mid \mathbf x_{t}, z_t=0)$.
On the other hand, we have $E\{\mu(\mathbf x_1^{t}, \mathbf z_1^{t-1},z_t)\mid \mathbf
x_{t}, z_t\}=\mu(\mathbf
x_{t},z_t)$ and $E\{\mu(\mathbf x_1^{t}, \mathbf z_1^{t-1},z_t=0)\mid \mathbf
x_{t}, z_t=0\}=\mu(\mathbf
x_{t},z_t=0)$.  Therefore, we have
$$E\{\theta(\mathbf{x}_{1}^{t}, \mathbf{z}_{1}^{t-1}; z_t)\mid \mathbf
x_{t}, z_t\}=\mu(\mathbf
x_{t},z_t)-\mu(\mathbf
x_{t},z_t=0),
$$
which is equal to  $\theta(\mathbf
x_{t}; z_t)$ according to (\ref{eq10.5}). This proves (\ref{eq10.5.1}).

{\bf Proof of formula (\ref{eq10.6.0})}:
Please notice that
for any function $f_j( \mathbf x_{s},z_s)$ of  $( \mathbf x_{s},z_s)$,
the expectation $E \{f_j( \mathbf x_{s},z_s)\mid \mathbf x_1^{t},\mathbf
z_1^{t}\}$ with respect to ${\rm P}(\mathbf{x}_{t+1}^{s},\mathbf z_{t+1}^{s}\mid \mathbf x_1^{t},\mathbf
z_1^{t})$ is equal to
$E \{f_j( \mathbf x_{s},z_s)\mid \mathbf x_1^{t},\mathbf
z_1^{t}\}$ with respect to ${\rm P}(\mathbf{x}_{s}, z_s\mid \mathbf x_1^{t},\mathbf
z_1^{t})$. Now inserting SNMM (\ref{eq10.3}) into the new $G$-formula (\ref{eq10.1}), we obtain
$$
\theta(\mathbf x_1^{t},\mathbf
z_1^{t-1};z_t)=\sum_{j=1}^k \gamma_j c_j(\mathbf x_1^{t},\mathbf
z_1^{t-1};z_t), \quad t=1,\dots, T,
$$
where
$$
c_j(\mathbf x_1^{t},\mathbf
z_1^{t-1};z_t)= f_j(\mathbf x_{t},z_t)
$$
$$
+\sum_{s=t+1}^T E_1 \{f_j( \mathbf x_{s},z_s)\mid \mathbf x_1^{t},\mathbf
z_1^{t-1},z_t\}
$$
$$
-\sum_{s=t+1}^T E_2 \{f_j(\mathbf x_{s} ,z_s)\mid \mathbf x_1^{t},\mathbf
z_1^{t-1},z_t=0\},
$$
where the conditional expectation $E_1(.)$ is with respect to    ${\rm P}(\mathbf{x}_{s}, z_s\mid \mathbf x_1^{t},\mathbf
z_1^{t-1},z_t)$ and $E_2(.)$ to ${\rm P}( \mathbf{x}_{s},z_s\mid \mathbf x_1^{t},\mathbf
z_1^{t-1},z_t=0)$.

Now, we average  both sides of the equality  with respect to ${\rm P}(\mathbf x_1^{t-1}, \mathbf z_1^{t-1}\mid \mathbf x_{t}, z_t)$.
According to  (\ref{eq10.5.1}), the average of the left side is equal to
$$
E\{\theta(\mathbf{x}_{1}^{t}, \mathbf{z}_{1}^{t-1}; z_t)\mid \mathbf
x_{t}, z_t\}=
\theta(\mathbf
x_{t}; z_t).
$$
The average of $E_1 \{f_j( \mathbf x_{s},z_s)\mid \mathbf x_1^{t},\mathbf
z_1^{t-1},z_t\}$  is equal to $E_1 \{f_j( \mathbf x_{s},z_s)\mid \mathbf x_{t}, z_t\}$.
 Due to (\ref{eq10.4}), the average of $E_2 \{f_j( \mathbf x_{s},z_s)\mid \mathbf x_1^{t},\mathbf
z_1^{t-1},z_t=0\}$  is equal to the average of
$E_2 \{f_j( \mathbf x_{s},z_s)\mid \mathbf x_1^{t},\mathbf
z_1^{t-1},z_t=0\}$ with respect to ${\rm P}(\mathbf x_1^{t-1}, \mathbf z_1^{t-1}\mid \mathbf x_{t}, z_t=0)$,
 which  is equal to
$E_2 \{f_j( \mathbf x_{s},z_s)\mid \mathbf x_{t}, z_t=0\}$. Putting these terms together, we obtain (\ref{eq10.6.0}).

{\bf Proof of formula (\ref{eq10.20})}: A special case of the point effect when $z_s$ is dichotomous   has been proved by Wang and Yin (2019). Here we extend the proof to  the point effect when $z_s$ is  discrete or continuous.
Conditional on $\{(\mathbf x_{i1}^{T},\mathbf z_{i1}^{T})\}$,  the point   effect  of treatment $z_s > 0$ in stratum $(\mathbf{x}_{1}^{s},\mathbf{z}_{1}^{s-1})$ is
$$
\theta_c(\mathbf{x}_{1}^{s},\mathbf{z}_{1}^{s-1}; z_s)=\mu_c(\mathbf{x}_{1}^{s},\mathbf{z}_{1}^{s-1},z_s)- \mu_c(\mathbf{x}_{1}^{s},\mathbf{z}_{1}^{s-1},z_s=0).
$$
We need the following lemma to prove (\ref{eq10.20})}.

\noindent {\bf Lemma}  {\textit{ Supposing the same  variance  $\sigma^2$ for the outcome $Y$ given any  $(
\mathbf x_{1}^{T}, \mathbf z_{1}^{T})$,  then the conditional covariance between the estimated point  effects at different times is equal to zero,  that is,
$$
{\rm cov}_c\{\hat\theta_c(\mathbf{x}_{1}^{s},\mathbf{z}_{1}^{s-1};z_s ); \hat\theta_c(\mathbf{x}_{1}^{t},\mathbf{z}_{1}^{t-1};z_t) \}=0, \quad s\neq t.
$$
}
\noindent Proof:
Without a loss of generality, we assume $s < t$.  Let $S$ be the stratum  of observations satisfying
$(\mathbf x_{i1}^{s},\mathbf z_{i1}^{s-1}, z_{is})=( \mathbf x_1^{s},\mathbf z_1^{s-1}, z_s)$; $S_0$   for $(\mathbf x_1^{t},\mathbf z_1^{t-1}, z_t=0)$;
$S_1$   for $(\mathbf x_1^{t},\mathbf z_1^{t-1}, z_t>0)$; and $S_2=S \ \backslash \ (S_1 \cup S_2)$.
Noticeably,  $S_0$ and $S_1$ are disjoint, and $S$ either contains both  $S_0$ and $S_1$ or contains neither. If $S$ contains neither  $S_0$ nor $S_1$, then the lemma is true. Therefore we only prove the lemma when $S$   contains both  $S_0$ and $S_1$.
Let $n(.)$ be the number of observations in a stratum. Then, $n(S)=n(S_0)+n(S_1)+n(S_2)$.
Rewrite
 $\mu_c(\mathbf x_1^{s},\mathbf z_1^{s-1}, z_s)=\mu(S)$, $\mu_c(\mathbf x_1^{t},\mathbf z_1^{t-1}, z_t=0)=\mu(S_0)$, $\mu_c( \mathbf x_1^{t},\mathbf z_1^{t-1}, z_t>0)=\mu(S_1)$.
Additionally, denote the mean of the outcome  $Y$ in $S_2$ by  $\mu(S_2)$.  Then, the mean $\mu(S)$ is equal to
$$
\mu(S)=\frac{n(S_0)}{n(S)}
\mu(S_0) +  \frac{n(S_1)}{n(S)}
\mu(S_1) +  \frac{n(S_2)}{n(S)}
\mu(S_2)
$$
and the  estimate of  $\mu(S)$ is
$$
\hat\mu(S)=\frac{n(S_0)}{n(S)}
\hat\mu(S_0) +  \frac{n(S_1)}{n(S)}
\hat\mu(S_1) +  \frac{n(S_2)}{n(S)}
\hat\mu(S_2).
$$
Thus,
$$
\hat\mu(S)-\mu(S)=\frac{n(S_0)}{n(S)}
\{\hat\mu(S_0)-\mu(S_0)\} +  \frac{n(S_1)}{n(S)}
\{\hat\mu(S_1)-\mu(S_1)\}
$$
$$+  \frac{n(S_2)}{n(S)}
\{\hat\mu(S_2)-\mu(S_2)\}.
$$
On the other hand, we have $\theta_c(\mathbf{x}_{1}^{t},\mathbf{z}_{1}^{t-1};z_t)=\mu(S_1)-\mu(S_0)$ and thus
$$
\hat \theta_c(\mathbf{x}_{1}^{t},\mathbf{z}_{1}^{t-1};z_t)-\theta_c(\mathbf{x}_{1}^{t},\mathbf{z}_{1}^{t-1};z_t)
=\{\hat\mu(S_1)-\mu(S_1)\}-\{\hat \mu(S_0)-\mu(S_0)\}.
$$
Recalling that $S_0$, $S_1$ and $S_2$ are disjoint, we have
 $$
 {\rm cov}_c\{\hat\mu(S), \hat\theta_c(\mathbf{x}_{1}^{t},\mathbf{z}_{1}^{t-1};z_t)\}
 $$
 $$=E_c\left[\{\hat\mu(S)-\mu(S)\}\{\hat \theta_c(\mathbf{x}_{1}^{t},\mathbf{z}_{1}^{t-1};z_t)-\theta_c( \mathbf{x}_{1}^{t}, \mathbf{z}_{1}^{t-1};z_t)\}
 \right]
 $$
 $$=
 \frac{n(S_1)}{n(S)}E_c\{\hat\mu(S_1)-\mu(S_1)\}^2 -\frac{n(S_0)}{n(S)}E_c\{\hat\mu(S_0)-\mu(S_0)\}^2,
 $$
 which is equal to
 $$
 \frac{\sigma^2}{n(S)}- \frac{\sigma^2}{n(S)}=0
 $$
 according to the assumption of
 the same  variance  $\sigma^2$ for $Y$ given any  $(
\mathbf x_{1}^{T}, \mathbf z_{1}^{T})$.
 Therefore,  we have
 $$
  {\rm cov}_c\{\hat\mu(S), \hat\theta_c(\mathbf{x}_{1}^{t},\mathbf{z}_{1}^{t-1};z_t)\} =0,
 $$
which is  true for all $\hat\mu(S)=\hat\mu_c(\mathbf x_1^{s},\mathbf z_1^{s-1},z_s)$. Noticeably,
 $\hat\theta_c(\mathbf{x}_{1}^{s},\mathbf{z}_{1}^{s-1};z_s)=\hat\mu_c(,\mathbf x_1^{s},\mathbf z_1^{s-1},$ $z_s)-\hat\mu_c( \mathbf x_1^{s},\mathbf z_1^{s-1} ,z_s=0)$; therefore,  we  have
$$
{\rm cov}_c\{ \hat\theta_c(\mathbf{x}_{1}^{s},\mathbf{z}_{1}^{s-1};z_s), \hat\theta_c(\mathbf{x}_{1}^{t},\mathbf{z}_{1}^{t-1};z_t )\}=0, \quad s<  t,
$$
which proves the lemma.

Now according to  (\ref{eq10.5.1}), conditional on  $\{(\mathbf x_{i1}^{T},\mathbf z_{i1}^{T})\}$,  we have
$$
\hat\theta_c(\mathbf
x_{s};z_s)=E_c\{\hat\theta_c(\mathbf{x}_{1}^{s}, \mathbf{z}_{1}^{s-1}; z_s)\mid \mathbf
x_{s}, z_s\},
$$
where the expectation is with respect to ${\rm P}_c(\mathbf x_1^{s-1}, \mathbf z_1^{s-1}\mid \mathbf x_{s}, z_s)$, and
$$
\hat\theta_c(\mathbf
x_{t};z_t)=E_c\{\hat\theta_c(\mathbf{x}_{1}^{t}, \mathbf{z}_{1}^{t-1}; z_t)\mid \mathbf
x_{t}, z_t\},
$$
where the expectation is with respect to ${\rm P}_c(\mathbf x_1^{t-1}, \mathbf z_1^{t-1}\mid \mathbf x_{t}, z_t)$.
These expressions  together with the lemma above imply
$$
{\rm cov}_c\{\hat\theta_c(\mathbf
x_{s};z_s);\hat\theta_c(\mathbf
x_{t};z_t)\}=0, \qquad s \neq t,
$$
which in turn implies
(\ref{eq10.20})}.

\pagebreak
\section*{Supplement II:  Simulation Study in Section $4$}
Here, we provide    details about the simulation study.
In Section $II.1$, we describe the testing procedure  in   the simulation study.
In Section $II.2$, we apply the procedure to the simulation.  In Section $II.3$, we construct the standard parameters, which generates the data.
The relevant SAS codes used for the simulation are   included in  Data and Codes of the article.

\subsection*{II.1 Testing procedure in  simulation}
In this simulation, SNMM  is known and so is the treatment assignment condition. Hence we start from the second stage of the testing procedure described in Section $3.5$.
In the second stage, we apply model (\ref{eq10.6.0})  to decompose the point effects into the blip effects
 and obtain
the following formulas (\ref{II_1}), (\ref{II_2}) and (\ref{II_3}):
\begin{subequations}
\begin{equation}\label{II_1}
\theta(z_1=1)=\gamma_1+\sum_{t=2}^3\sum_{ i=0}^3\gamma_{ti}c_{ti},
\end{equation}
where $c_{ti}={\rm P}(x_{t}=i, z_t=1\mid z_1=1)-
 {\rm P}( x_{t}=i, z_t=1\mid z_1=0) $;
\begin{equation}\label{II_2}
 \theta( x_2=j;z_2=1)=\gamma_{2j}+\sum_{ i=0}^3\gamma_{3i}c_{3i}(x_2=j),
\end{equation}
where  $c_{3i}(x_2=j)={\rm P}(x_{3}=i, z_3=1\mid x_2=j,z_2=1)-
 {\rm P}( x_{3}=i, z_3=1\mid x_2=j,z_2=0)$;
\begin{equation}\label{II_3}
\theta( x_3=j; z_3=1)=\gamma_{3j}.
\end{equation}
\end{subequations}

In the third stage, we  estimate the   blip effect parameter    $\boldsymbol\gamma=(\gamma_1, \gamma_{20},$ $\gamma_{21}, \gamma_{22}, \gamma_{23},
\gamma_{30}, \gamma_{31}, \gamma_{32}, \gamma_{33})'$.
Let $n(x_{t}, z_t)$ be the number of observations  in stratum $(x_{t}, z_t)$,  and $I(x_{t}, z_t)$  be the set of
  all indexes $i$ such that $
(x_{it},z_{it})=
(x_{t},z_t)$. For the normal outcome,
 we have
 $$
\hat\theta_c(
x_{t};z_t=1)=
\frac{\sum_{i\in I(x_{t}, z_t=1)} y_i}{n(
x_{t},z_t=1)}-\frac{\sum_{i\in I(x_{t}, z_t=0)} y_i}{n(
x_{t},z_t=0)},
$$
$$
{\rm var}_c\{\hat\theta_c(
x_{t};z_t=1)\}={\sigma^2 \over n(
x_{t},z_t=1)}+{\sigma^2 \over n(
x_{t},z_t=0)},
$$
where $\sigma^2$ is the conditional variance of $Y$ given $(z_1, x_2, z_2, x_3, z_3)$.
 For the  dichotomous and Poisson  outcomes, we use standard softwares to obtain $\hat\theta_c(
x_{t};z_t=1)$ and ${\rm var}_c\{\hat\theta_c(
x_{t};z_t=1)\}$.

The probabilities in (\ref{II_1}), (\ref{II_2}) and (\ref{II_3}) are estimated by the corresponding proportions, which in turn lead to evaluation of  $c_{c,2i}$, $ c_{c,3i}$, and $c_{c,3i}(x_2)$, $i=0,1,2,3$.
We estimate ${\boldsymbol \gamma}_c $ by regressing the obtained  $\hat\theta_c(z_1=1)$ on one, $ c_{c,2i}$ and $c_{c,3i}$ based on  (\ref{II_1}); $\hat\theta_c(
x_2;z_2=1)$ on one  and $ c_{c,3i}(x_2)$ based on  (\ref{II_2}); and $\hat\theta_c(
x_3;z_3=1)$ on one based on  (\ref{II_3}). According to Proposition \ref{P2}, the obtained conditional estimate $\widehat{\boldsymbol\gamma}_c$   is  equal to  the marginal estimate $\widehat{\boldsymbol\gamma}$.

In the  fourth stage, we estimate  the marginal  covariance matrix ${\rm cov}(\widehat {\boldsymbol\gamma})$ by applying the bootstrap method to the third stage above.
With ${\rm cov}(\widehat {\boldsymbol\gamma})$ as well as $\widehat{\boldsymbol\gamma}$, we apply (\ref{eq30.2})  to calculate  the Wald statistics for  hypothesis $A$ through $J$. With the obtained Wald statistic, we
 test the hypothesis at the significance  level $\alpha=0.05$ according to Theorem \ref{T1}.

\subsection*{II.2 Application of  testing procedure  to  simulation}
 Three data-generating
mechanisms are constructed for normal,  dichotomous and Poisson outcomes  using the
standard parameters obtained in the next section.
The sample sizes are chosen as  $1000$, $2000$ and $3000$, such that the point effects are estimable.
For every sample size,  $1000$ data sets are generated   of each outcome type.

We apply the procedure described in the previous subsection to each of these $1000$ data sets and
check if the type I and II errors occur for hypothesis  $A$ through $J$.
The covariance matrix  for the  blip effects is estimated by using the bootstrap method with only $500$ replications  due to our limited computing power.

From these $1000$ data sets, we obtain the type I and II error rates, which are presented in Table $1$. In Table $2$, we   present the estimate and its variance for  the blip effect under no constraint or under the null hypothesis $J_0$: $(\gamma_{20},\gamma_{21}, \gamma_{22}, \gamma_{23})=(\gamma_{30},\gamma_{31}, \gamma_{32}, \gamma_{33})$.

\subsection*{II.3 Construction of  standard parameters for  normal, dichotomous and Poisson outcomes}
The probabilities of   treatments and covariates are the same for the three outcome types and presented in Table $II1$.
Here, we will use the point   effects of treatments,  the point   effects of covariates and the grand mean to construct the standard parameters for the conditional distribution of the outcome given all   treatments and covariates. These standard parameters yield true values of the blip effects in the simulation (Wang and Yin, 2015).

As described in Section $4$, we have the following blip effects.
At $t=1$, there is only one blip effect of $z_1=1$:  $\phi(z_1=1)   =\gamma_1$. At $t=2$, there are  four blip effects  of $z_2=1$ depending only on $x_2=0,1,2,3$:
$\phi(z_1,   x_2 =j;z_1=1 )=\gamma_{2j}$,   $j=0,1,2,3$. At $t=3$, there are  four blip effects  of $z_3=1$ depending only on $ x_3=0,1,2,3$:
$\phi(z_1,   x_2, z_2, x_3=j;z_3=1  )=\gamma_{3j}$,  $j=0,1,2,3$.
Inserting these  blip   effects into the new $G$-formula (\ref{eq10.1}), we obtain the formula for calculating the point   effect of treatment
$$
\left\{
\begin{array}{l}
\theta(z_1=1)=\gamma_1+\sum_{t=2}^3\sum_{i=0}^3\gamma_{ti} {\rm P} ( x_{t}=i, z_t=1\mid z_1=1)
\\
\qquad\qquad\qquad    -\sum_{t=2}^3\sum_{i=0}^3\gamma_{ti} {\rm P} ( x_{t}=i, z_t=1\mid z_1=0),
\\
\theta(z_1, x_2=j;z_2=1)=\gamma_{2j}+\sum_{i=0}^3 \gamma_{3i}{\rm P} ( x_3=i, z_3=1\mid z_1,  x_2=j, z_2=1)
\\
\qquad\qquad\qquad\qquad\qquad -\sum_{i=0}^3\gamma_{3i}{\rm P} ( x_3=i, z_3=1\mid z_1,  x_2=j, z_2=0),

\\
\theta(z_1,  x_2, z_2,  x_3=j;z_3=1)=\gamma_{3j}.
\end{array}
\right.
$$
Using the  true values of the  blip  effects    and  the probabilities of   treatments and covariates   given in Table $II1$, we  calculate these point   effects of the treatments.

The point   effect of covariate $x_{2} > 0$ is $\zeta( z_1;  x_{2})=\mu( z_1, x_{2})-\mu( z_1,  x_{2}=0)$, where
$\mu( z_1, x_{2})=E(Y\mid z_1, x_2)$.
The point   effect of covariate $x_{3} > 0$ is $\zeta( z_1, x_2, z_2;  x_{3})=\mu( z_1, x_2, z_2,  x_{3})-\mu( z_1, x_2, z_2,  x_{3}=0)$, where
$\mu( z_1, x_2, z_2,  x_{3})=E(Y\mid  z_1, x_2, z_2,  x_{3})$.
The grand mean is $\mu=E(Y)$. According to the new $G$-formula (\ref{eq10.1}) or its inverse form (Wang and Yin, 2019), the  blip effects  are only functions of the point effects of treatments; therefore,
 the point   effects  $\zeta( z_1;  x_{2})$ and $\zeta( z_1, x_2, z_2;  x_{3})$ and the grand mean  can be arbitrarily chosen to yield the same blip effects. However, the choice should  allow for an appropriate mean of the distribution, e.g., the mean must have a range of $(0,1)$ for a dichotomous outcome.

For the normal distribution, we choose the   point effects of covariates
$$
\zeta(z_1; x_2) =\left\{ \begin{array}{l} 10+ 5z_1, \quad  x_2=1
\\
12+ 5z_1, \quad  x_2=2
\\
13+ 5z_1, \quad   x_2=3
\end{array} \right.
$$
for $z_1=0,1$, and
$$
\zeta(z_1, x_2, z_2;  x_3) =\left\{ \begin{array}{l} 10- 5z_1 -2z_2+3x_2, \quad   x_3=1
\\
12- 5z_1 -2z_2+3x_2, \quad  x_3=2
\\
10- 5z_1 -3z_2+3x_2, \quad  x_3=3
\end{array} \right.
$$
for $z_1=0,1$, $z_2=0,1$ and $ x_2=0,1,2,3$ respectively.
We choose the grand mean as $\mu=-5$.

For the dichotomous outcome, we choose the  point effects of covariates
$$
\zeta(z_1;x_2) =\left\{ \begin{array}{l}  0.1z_1, \quad   x_2=1
\\
0.1 z_1, \quad   x_2=2
\\
0.1 z_1, \quad   x_2=3
\end{array} \right.
$$
for $z_1=0,1$, and
$$
\zeta(z_1,  x_2, z_2;   x_3) =\left\{ \begin{array}{l}  -0.1z_2, \quad   x_3=1
\\
  -0.1z_2 , \quad   x_3=2
\\
-0.1z_2, \quad  x_3=3
\end{array} \right.
$$
for $z_1=0,1$, $z_2=0,1$ and   $ x_2=0,1,2,3$.
We choose  the grand mean as $\mu=0.55$.

For the Poisson  outcome, we choose the same point effects of covariates   as those for the dichotomous  outcome.
As for the grand mean, we choose  $\mu=20$.

Finally, we use the obtained  $\theta(\mathbf z_1^{t-1},\mathbf x_2^{t}; z_t)$, $\zeta(\mathbf z_1^{t-1},\mathbf x_2^{t-1}; x_{t})$ and $\mu$ to construct the standard parameter $\mu(z_1,  x_2, z_2,  x_3, z_3)$ by applying  formula (16) of Wang and Yin (2015),  that is,
$$
\mu(z_1, x_2, z_2, x_3, z_3)=-\sum_{t=1}^3 \theta(\mathbf z_1^{t-1},\mathbf x_2^{t}; z_t^*=1)
\{{\rm P}(z_t^*=1\mid \mathbf z_1^{t-1},\mathbf x_2^{t})-I(z_t)\}
$$
$$
-\sum_{t=2}^{3} \left \{\sum_{  x_{t}^* > 0}
\zeta(\mathbf z_1^{t-1},\mathbf x_2^{t-1};   x_{t}^*){\rm P}(  x_{t}^*\mid \mathbf z_1^{t-1},\mathbf x_2^{t-1})-
\zeta(\mathbf z_1^{t-1},\mathbf x_2^{t-1};  x_{t})\right \}+
\mu,
$$
where $I(z_t)$ equals one when $z_t=1$ and zero otherwise.  The obtained standard parameters are presented in Tables $II2$-$II4$ for the normal, Bernoulli and Poisson   distributions respectively.

\end{document}